\documentclass[useAMS,usenatbib]{mn2e}

\usepackage{graphicx}

\newcommand{\beq}{\begin{equation}}
\newcommand{\eeq}{\end{equation}}
\newcommand{\lab}{\label}

\newcommand{\bfxi}{\mbox{\boldmath $\xi$}}
\newcommand{\bfeta}{\mbox{\boldmath $\eta$}}
\newcommand{\bfomega}{\mbox{\boldmath $\omega$}}
\newcommand{\bftheta}{\mbox{\boldmath $\theta$}}
\newcommand{\bfal}{\mbox{\boldmath $\alpha$}}
\newcommand{\bfx}{{\bf x}}
\newcommand{\bfy}{{\bf y}}

\begin{document}

\title[On gravito-magnetic time delay]{On gravito-magnetic time delay by extended lenses}
\author[M. Sereno]{M. Sereno$^{1,2,3}$\thanks{E-mail:
mauro.sereno@na.infn.it}
\\
$^{1}$Dipartimento di Scienze Fisiche, Universit\`{a} degli Studi di
Napoli ``Federico II", Via Cinthia, Monte S. Angelo, 80126 Napoli,
Italia
\\
$^{2}$Istituto Nazionale di Astrofisica - Osservatorio Astronomico di
Capodimonte, Salita Moiariello, 16, 80131 Napoli, Italia
\\
$^{3}$Istituto Nazionale di Fisica Nucleare, Sez. Napoli, Via Cinthia,
Monte S. Angelo, 80126 Napoli, Italia}

\date{13 September, 2004}

\maketitle

\begin{abstract}
Gravitational lensing by rotating extended deflectors is discussed.
Due to spin, corrections on image positions, caustics and critical
curves can be significant. In order to obtain realistic quantitative
estimates, the lens is modeled as a singular isothermal sphere.
Gravito-magnetic time delays of $\sim 0.2$~days between different
images of background sources can occur.
\end{abstract}

\begin{keywords}
astrometry -- gravitation -- gravitational lensing -- relativity
\end{keywords}

\section{Introduction}

Mass-energy currents relative to other masses generate space-time
curvature. This phenomenon, known as intrinsic gravito-magnetism, is a
new feature of general theory of relativity and other conceivable
post-Newtonian theories of gravity \citep[see][and references
therein]{ci+wh95}. The gravito-magnetic field has not been yet
detected with high accuracy. Some results have been reported from
laser-ranged satellites, when the Lense-Thirring precession, due to
the Earth spin, was measured by studying the orbital perturbations of
LAGEOS and LAGEOS II satellites \citep{ci+pa98,ci+pa04}. According to
a preliminary analysis, the predictions of the general theory of
relativity have been found to agree with the experimental values
within the $\sim 10$ per cent accuracy \citep{ci+pa04}. The NASA's
Gravity Probe B satellite should improve this measurement to an
accuracy of 1 per cent. Intrinsic gravito-magnetism might play a
relevant role also in the dynamics of the accretion disk of a
supermassive black hole or in the alignment of jets in active galactic
nuclei and quasars \citep{ci+wh95}.

Whereas the tests just mentioned limit to the gravitational field
outside a spinning body, the general theory of relativity predicts
peculiar phenomena also inside a rotating shell \citep[see][for
example]{wei72}. Gravitational lensing can represent a tool to fully
test the effects of the gravito-magnetic field (see \citet{ser03} and
reference therein). In this paper, we are interested in the
gravito-magnetic time delay induced in different images of the same
source due to gravitational lensing. Whereas here we are mainly faced
with intrinsic gravito-magnetism, i.e. with spinning deflectors, a
translational motion of the lens can also induce interesting
phenomena, which in the framework of general relativity are strictly
connected to the Lorentz transformation properties of the
gravitational field. The effect of the deflector's velocity has been
recently observed in the Jovian deflection experiment conducted at
VLBI, which measured the time delay of light from a background quasar
\citep{fo+ko03}. Although a controversy has emerged over the
theoretical interpretation of this measurement \citep{kop04}, there is
agreement about the role of the deflector's motion.

Gravitational time delay by spinning deflectors has been addressed by
several authors with very different approaches. \citet{dym86}
discussed the additional time-delay due to rotation by integrating the
light geodesics of the Kerr metric. Using the Lense-Thirring metric,
\citet{gli99} considered the time delay for light rays passing outside
a spinning star. Kopeikin and collaborators
\citep{kop97,ko+sc99,ko+ma02} analysed the gravito-magnetic effects in
the propagation of light in the field of self-gravitating spinning
bodies. The gravitational time delay due to rotating masses was
further discussed in \citet{ciu+al03,ci+ri03}, where the cases of
light rays crossing a slowly rotating shell or propagating in the
field of a distant source were analyzed in the linear approximation of
general theory of relativity. Effects of an intrinsic gravito-magnetic
field were further studied in the usual framework of gravitational
lensing theory \citep{ser02pla,ser03prd,ser03,se+ca02}, i.e. {\it i)}
weak field and slow motion approximation for the lens and {\it ii)}
thin lens hypothesis \citep{sef,pet+al01}. Expressions for bending and
time delay of electromagnetic waves were found for stationary spinning
deflectors with general mass distributions \citep{ser02pla}.

The paper is as follows. In Section~\ref{basi}, basics of
gravitational lensing by a stationary extended deflector are reviewed.
In Section~\ref{sis}, we introduce our reference model for the lens,
i.e. a spinning singular isothermal sphere. Relevant lensing
quantities for such a deflecting system are derived in
Section~\ref{lensBySIS}. In Section~\ref{lens}, the lens equation is
solved. A quantitative discussion of the gravito-magnetic time delay
is in Section~\ref{dela}, where we also investigate the effect on the
determination of the Hubble constant. Section~\ref{disc} is devoted to
some final considerations. Unless otherwise stated, throughout this
paper we consider a flat cosmological model of universe with
cosmological constant and a pressureless cosmological density
parameter $\Omega_{\rm M0}=0.3$ a Hubble constant
$H_0=72~\mathrm{km~s}^{-1}\mathrm{Mpc}^{-1}$ .

\section{Gravito-magnetic deflection potential}
\label{basi}

Gravitational lensing theory can be easily developed in the
gravitational field of a rotating stationary source when the linear
approximation of general relativity holds \citep{ser02pla}. The time
delay of a kinematically possible ray, with impact parameter $\bfxi$
in the lens plane, relative to the unlensed one is, for a single lens
plane,\citep{ser02pla}
\beq
\lab{pot1}
\Delta t =\frac{(1+z_{\rm d})}{c}\left\{ \frac{1}{2}\frac{D_\mathrm{d}
D_\mathrm{s} }{D_\mathrm{ds}}\left|
\frac{\bfxi}{ D_\mathrm{d} }-\frac{\bfeta}{ D_\mathrm{s} } \right|^2 -
\hat{\psi}(\bfxi ) \right\},
\eeq
where $\hat{\psi}$ is the deflection potential; $D_\mathrm{s}$,
$D_{\rm d}$ and $D_{\rm ds}$ are the angular diameter distances
between observer and source, observer and lens and lens and source,
respectively; $z_{\rm d}$ is the lens redshift; $\bfeta$ is the
bidimensional vector position of the source in the source plane. We
have neglected a constant term in equation~(\ref{pot1}), since it has
no physical significance \citep{sef}.

The deflection potential can be expressed as the sum of two terms
\beq
\hat{\psi} \simeq \hat{\psi}_0+ \hat{\psi}_\mathrm{GRM};
\eeq
the main contribution is
\beq
\lab{pot2}
\hat{\psi}_0(\bfxi ) \equiv \frac{4 G}{c^2}\int_{\Re^2}d^2\xi^{'}\Sigma(\bfxi^{'})
\ln \frac{|\bfxi -\bfxi^{'}|}{\xi_0},
\eeq
where $\xi_0$ is a length scale in the lens plane and $\Sigma$ is the
projected surface mass density of the deflector,
\beq
\lab{pot3}
\Sigma(\bfxi)\equiv \int \rho(\bfxi,l)\ dl;
\eeq
the gravito-magnetic correction to the deflection potential, up to the
order $v/c$, can be expressed as \citep{ser02pla}
\beq
\hat{\psi}_\mathrm{GRM} \simeq -
\frac{8 G}{c^4} \int_{\Re^2} d^2 \xi^{'} \Sigma(\bfxi^{'})
\langle {\bf v}{\cdot}{\bf e}_\mathrm{in} \rangle_l (\bfxi^{'})\ln
\frac{|\bfxi -\bfxi^{'}|}{\xi_0},
\eeq
where $\langle {\bf v}{\cdot}{\bf e}_\mathrm{in}\rangle_l$ is the weighted
average, along the line of sight ${\bf e}_\mathrm{in}$, of the
component of the velocity $\bf v$ along ${\bf e}_\mathrm{in}$,
\beq
\lab{pot4}
\langle {\bf v}{\cdot}{\bf e}_\mathrm{in}\rangle_l (\bfxi)\equiv \frac{\int ({\bf v}(\bfxi,l){\cdot} {\bf e}_\mathrm{in})
\ \rho(\bfxi,l)\ dl}{\Sigma(\bfxi)}.
\eeq
In the thin lens approximation, the only components of the velocities
parallel to the line of sight enter the equations of gravitational
lensing. We remind that the time delay function is not an observable,
but the time delay between two actual rays can be measured. Similar
results, based on a multi-polar description of the gravitational field
of a stationary lens, can be found in \citet{kop97}.

\section{Singular isothermal sphere}
\label{sis}

Isothermal spheres are widely used in astrophysics to understand many
properties of systems on very different scales, from galaxy haloes to
clusters of galaxies \citep{mo+al98,sef}. In particular, on the scale
relevant to interpreting time delays, isothermal models are favoured
by data on early-type galaxies \citep{ko+sc04}. The density profile of
a singular isothermal sphere (SIS) is
\beq
\lab{sis7}
\rho (r) = \frac{\sigma_\mathrm{v}^2}{2 \pi G r^2},
\eeq
where $\sigma_\mathrm{v}$ is the velocity dispersion. The
corresponding projected mass density is
\beq
\lab{sis9}
\Sigma(\xi)= \frac{\sigma_\mathrm{v} ^2}{2G} \frac{1}{\xi}.
\eeq

Since the total mass is divergent, we introduce a cut-off radius
$R_\mathrm{SIS} \gg \xi$. The cut-off radius of the halo must be much
larger than the relevant length scale which characterizes the
phenomenon, in order to not significantly affect the lensing behavior.
The total mass of a truncated SIS is
\beq
M_\mathrm{SIS}=\frac{2 \sigma_\mathrm{v}^2}{G} R_\mathrm{SIS}.
\eeq
A limiting radius can be defined as $r_n$, the radius within which the
mean mass density is $n$ times the critical density of the universe at
the redshift of the galaxy, $z_{\rm d}$. For a SIS, it is
\beq
r_n = \frac{2 \sigma_\mathrm{v} }{ \sqrt{n} H(z_{\rm d})},
\eeq
where $H$ is the time dependent Hubble parameter.

The total angular momentum of an halo, $J$, can be expressed in terms
of a dimensionless spin parameter $\lambda$, which represents the
ratio between the actual angular velocity of the system and the
hypothetical angular velocity that is needed to support the system
\citep{pad02},
\beq
J  \equiv \lambda \frac{G M^{5/2}}{|E|^{1/2}},
\eeq
where $M$ and $E$ are the total mass and the total energy of the halo,
respectively. In the hypothesis of initial angular momentum acquired
from tidal torquing, typical values of $\lambda$ can be obtained from
the relation between energy and virial radius and the details of
spherical top-hat model \citep{pad02}. As it was derived from
numerical simulations, the distribution of $\lambda$ is nearly
independent of the mass and the power spectrum. It can be approximated
by a log-normal distribution \citep{vit+al02}
\beq
p(\lambda) d \lambda = \frac{1}{\sqrt{2\pi}\sigma_\lambda}\exp \left[
-\frac{\ln^2 (\lambda/\bar{\lambda})}{2\sigma_\lambda^2} \right]
\frac{d \lambda}{\lambda},
\eeq
with $\bar{\lambda} \simeq 0.05$ and $\sigma_\lambda \simeq 0.5$. The
distribution peaks around $\lambda \simeq 0.04$ and has a width of
$\sim 0.05$.

From the virial theorem, the total energy is easily obtained
\citep{mo+al98}
\beq
E_\mathrm{SIS} = - M_\mathrm{SIS} \sigma_\mathrm{v}^2.
\eeq
Finally, the total angular momentum of a truncated SIS can be written
as
\beq
J_\mathrm{SIS}= \lambda \frac{4 \sigma_\mathrm{v}^3
R_\mathrm{SIS}^2}{G}.
\eeq
In general, the angular velocity of a halo is not constant and a
differential rotation should be considered \citep{cap+al03}. However,
assuming a detailed rotation pattern does not affect significantly the
results. In what follows, we will consider the case of constant
angular velocity.

\section{Lensing by a rotating SIS}
\label{lensBySIS}

Let us consider gravitational lensing by a SIS in rigid rotation, i.e
with a constant angular velocity $\bfomega$, about an arbitrary axis
passing through its center. The deflection angle can be written as
\citep{se+ca02}
\begin{eqnarray}
\hat{\alpha}_1^{\rm SIS}(\xi, \vartheta) & =& 4 \pi
\left( \frac{\sigma_\mathrm{v}}{c} \right)^2 \left\{   \cos \vartheta  +
\frac{\omega_1}{c}\xi  \frac{\sin 2 \vartheta}{3}
\right. \nonumber
\\
& & - \left. \frac{\omega_2}{c}\left[ \xi \left( \frac{\cos 2
\vartheta}{3} +1 \right)
-R_\mathrm{SIS}  \right]
\right\}, \label{def1}
\\
\hat{\alpha}_2^{\rm SIS}(\xi, \vartheta) & =& 4 \pi \left( \frac{\sigma_\mathrm{v}}{c}
\right)^2 \left\{   \sin \theta  +
\frac{\omega_2}{c}\xi  \frac{\sin 2 \vartheta}{3}
\right. \nonumber \\
&  & +\left. \frac{\omega_1}{c}\left[ \xi \left( \frac{\cos 2
\theta}{3} -1 \right) + R_\mathrm{SIS}  \right]
\right\} , \label{def2}
\end{eqnarray}
where $(\xi,\vartheta)$ are polar coordinates in the lens plane and
$\omega_1$ and $\omega_2$ are the components of $\bfomega$ along the
$\xi_1$- and the $\xi_2$-axis, respectively. The parameter $\omega$
has to be interpreted as an effective angular velocity $\tilde{\omega}
\equiv J_{\rm SIS}/I_{\rm SIS}$ where $I_{\rm SIS}$ is the central
momentum of inertia of a truncated SIS, $I_{\rm SIS} = 2/9 M_{\rm SIS}
R_\mathrm{SIS}^2$. In terms of the spin parameter,
\beq
\tilde{\omega} = 9 \lambda \frac{\sigma_\mathrm{v}}{R_\mathrm{SIS}}.
\eeq
Equations~(\ref{def1}) and (\ref{def2}) correct the result in
equation~(9) in \citet{ca+re01}, which was obtained under the same
assumptions but was affected by an error in the computation of the
integrals. For a SIS, there are two main contributions to the
gravito-magnetic correction to the deflection angle \citep{se+ca02}:
the first contribution comes from the projected momentum of inertia
inside the radius $\xi$; the second contribution is due to the mass
outside $\xi$ and can become significant in the case of a very
extended lens, i.e. for a very large cut-off radius.

Let us consider a sphere rotating about the $\xi_2$-axis,
$\omega_1=0,\omega_2 =\omega$. In order to change to dimensionless
variables, we introduce a length scale,
\beq
\lab{sis10}
\xi_0 =R_{\rm E}= 4 \pi \left( \frac{\sigma_\mathrm{v}}{c}\right) ^2
\frac{D_{\rm d} D_{\rm ds}}{D_{\rm s}}.
\eeq
The dimensionless position vector in the lens plane is $\bfx \equiv
\bfxi/\xi_0$. The dimensionless deflection
potential $\psi$,
\beq
\lab{eq8}
\psi \equiv \frac{D_{\rm d} D_{\rm ds}}{D_{\rm s} \xi _0 ^2} \hat{\psi}
\eeq
can be written as
\beq
\psi^{\rm SIS}(x_1,x_2)=x -(\frac{3}{2}r-x) L x_1,
\eeq
where $x \equiv |\bfx|$ and $L \equiv (2/3)(\tilde{\omega}
R_\mathrm{E}/c)$ is an estimate of the rotational velocity and $r$ is
the cut-off radius in units of $R_{\rm E}$. When $L>0$, the angular
momentum of the lens is positively oriented along $\hat{x}_2$. The
dimensionless Fermat potential $\phi$ is defined as,
\beq
\phi (\bfx , \bfy ) = \frac{1}{2} (\bfx - \bfy )^2 -\psi (\bfx ) \lab{eq10},
\eeq
where $\bfy \equiv \bfeta/\left( \frac{D_{\rm s}}{D_{\rm d}}\xi_0
\right)$.

The scaled deflection angle is related to the dimensionless
gravitational potential $\psi$
\beq
\bfal (\bfx) = \nabla \psi (\bfx) .
\eeq
We get
\begin{eqnarray}
\alpha_1^{\rm SIS}(x_1,x_2) & = & \frac{x_1}{x}+L\left( \frac{2x_1^2 +x_2^2}{x}-\frac{3}{2}r \right), \\
\alpha_2^{\rm SIS}(x_1,x_2) & = & \frac{x_2}{x}+L\frac{x_1 x_2}{x} .
\end{eqnarray}
The determinant of the Jacobian matrix reads
\beq
A_{\rm SIS}(x_1,x_2) \simeq 1-\frac{1}{x}-L\frac{x_1}{x}\left(
3-\frac{2}{x}
\right) .
\eeq
The convergence $k \equiv \Sigma/\Sigma_{\rm cr}$ is the projected
density in units of the critical surface density,
\beq
\Sigma_{\rm cr} \equiv \frac{c^2}{4 \pi G} \frac{D_{\rm s}}{D_{\rm d} D_{\rm ds}}.
\eeq
It is
\beq
k_{\rm SIS}=\frac{1+3Lx_1}{2 x};
\eeq
$k_{\rm SIS}$ is positive when $x_1 > -1/3L$. Since we have introduced
a cut-off radius, this condition holds for $x_1
\stackrel{<}{\sim} r$. The condition $L r < 1/3$ warrants that
$k_{\rm SIS}>0$ for all points in the lens plane.

\section{The lens equation}
\label{lens}

\begin{figure}
        \resizebox{\hsize}{!}{\includegraphics{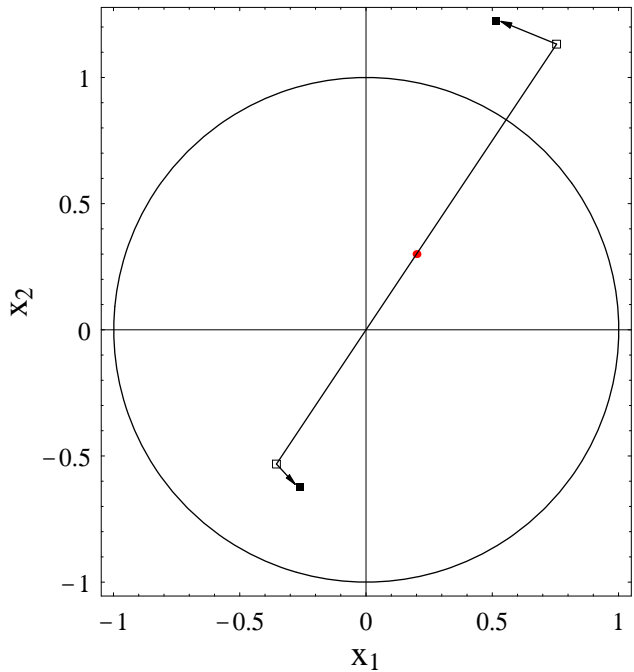}}
        \caption{Image positions (square boxes) and critical line (full line) for a rotating SIS lens.
        The source is the grey circle. The centre of the coordinate-axes, the source
        and the two unperturbed images (empty boxes) lie on a straight line. Two images (filled boxes)
        are counter-clockwisely rotated, about the line of sight through the centre,
        with respect to this line. It is $L=10^{-3}$ and $r=50$.}
        \label{SISImagesArrows}
\end{figure}

\begin{figure}
        \resizebox{\hsize}{!}{\includegraphics{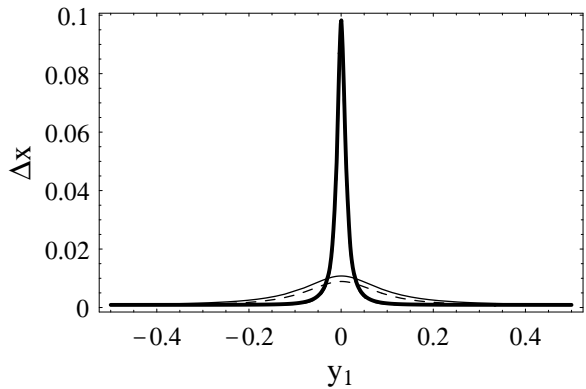}}
        \caption{The shift in the image positions (in units of arcsec) due to the
        gravito-magnetic field for source moving parallelly to the $y_1$-axis. Thick and thin lines
are for
        $y_2=0.01$ and 0.1, respectively.
        Full and dashed lines refer to the two images. It is $L=4{\times} 10^{-5}$ and $r=15$.}
        \label{SISDeltaXGRM}
\end{figure}

\begin{figure}
        \resizebox{\hsize}{!}{\includegraphics{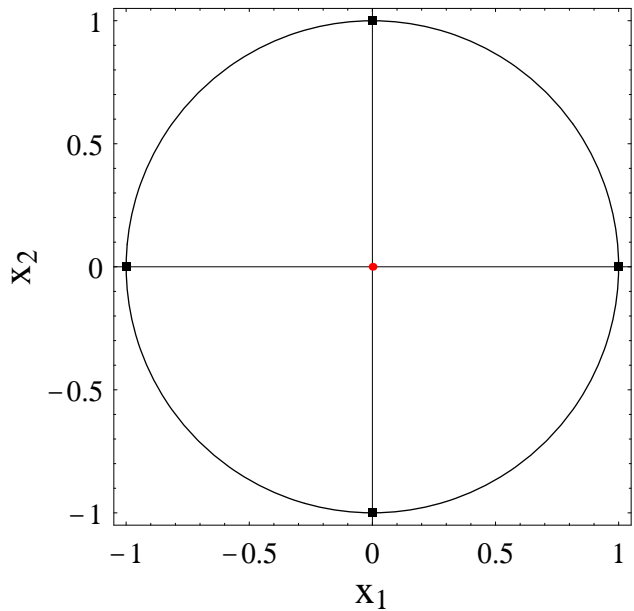}}
        \caption{A source (the grey circle) inside the central caustic of a rotating
        SIS is multiple imaged in a cross shaped pattern; the four filled box locate
        the four images. The critical line is also plotted. It is $r=15$ and
        $L=3.5 {\times} 10^{-5}$.}
        \label{SISCrossGRM}
\end{figure}

In general, the inversion of the lens equation is a mathematical
demanding problem. However, since the gravito-magnetic effect in the
weak field limit is an higher-order correction, a rotating system can
be studied in some details using a perturbative approach
\citep{ser03}. The lens equation can be expressed in a dimensionless
form as
\beq
\label{lens1}
\bfy = \bfx - \bfal (\bfx).
\eeq
The unperturbed images are solutions of the lens equation for $L=0$.
When $ y \equiv |\bfy | < 1$, a static SIS lens produces two images,
collinear with the source position and the lens centre, at radii $x=
y+1$ and $x= y-1$. When $y
>1$, only one image forms, at $x= y+1$.

Under the condition $L \ll 1$, we can obtain approximate solutions to
the first-order in $L$, given by
\beq
\label{sost}
\bfx \simeq \bfx_{(0)}+L\bfx_{(1)},
\eeq
where $\bfx_{(0)}$ and $\bfx_{(1)}$ denote the zeroth-order solution
(i.e. the solution of the lens equation for $L=0$), and the correction
to the first-order, respectively. By substituting
equation~(\ref{sost}) in the vectorial lens equation,
equation~(\ref{lens1}), we obtain the first-order perturbations,
\begin{eqnarray}
x_{(1)1}& =&  x_{(0)}^2+\left( \frac{3}{2}r-2x_{(0)} +x_{(0)}^2
\right)\frac{x_{(0)1}^2-x_{(0)}^3}{x_{(0)}^2(x_{(0)}-1)}, \lab{def3}\\
x_{(1)2}& =&  \left( \frac{3}{2}r-2x_{(0)} +x_{(0)}^2 \right)
\frac{x_{(0)1} x_{(0)2} }{x_{(0)}^2(x_{(0)}-1)}. \lab{def4}
\end{eqnarray}
For a source on the $y_1$-axis ($y_2=0$), equations~(\ref{def3}) and
(\ref{def4}) reduce to
\begin{eqnarray}
x_{(1)1}& = &   - \frac{3}{2}r + 2|x_{(0)1}|
\\
x_{(1)2}& = & 0 .
\end{eqnarray}
When $L>0$, photons with an impact parameter $x_1<0$ ($x_1>0$) go
around the lens in the same (opposite) sense of the deflector. Due to
the contribution to the deflection angle from the mass outside the
impact parameter, photons which impact the lens plane on the
$x_1$-axis at $x_1 > 0$ move closer to the centre. This feature, which
is peculiar of light rays propagating inside a halo, is opposite to
the case of propagation outside a spinning sphere \citep{ser03}, when
photons in the equatorial plane moving in the same (opposite) sense of
rotation of a deflector form closer (farther) images with respect to
the non-rotating case. In fact, in the last case, there is no mass
outside the photon path. If we do not limit to the equatorial plane,
for $L>0$, the images are rotated counter-clockwisely around the line
of sight with respect to the static case, see
Fig.~\ref{SISImagesArrows}.

Let us consider the gravito-magnetic correction to the deflection
angle for a typical galaxy lens at $z_{\rm d}=0.3$  with $\sigma_{\rm
v} \sim 250$~Km~s$^{-1}$, $R_\mathrm{SIS} \stackrel{<}{\sim} 100$~Kpc
and $\lambda \sim 0.1$, and a background quasar at $z_{\rm s}=2.0$.
Such a configuration correspond to $L \sim (2-4) {\times} 10^{-5}$ and
$r\simeq 15$. For some particular source positions, the shift in the
image positions in the lens plane with respect to the static case can
be as large as $0.1\arcsec$. In Fig.~\ref{SISDeltaXGRM}, we plot the
shift in the image positions for a source moving along the $y_1$-axis,
i.e for $y_2 = const.$ The maximum variation occurs when the source
nearly crosses the projected rotation axis.

The critical curve is slightly distorted. The solution of $\det
A(x_1,x_2)=0$, with respect to $x_2$, is
\begin{equation}
x_2(x_1) \simeq \pm \left\{
\sqrt{1-x_1^2}+\frac{x_1}{\sqrt{1-x_1^2}}L
\right\}.
\end{equation}
The area of the critical curve slightly grows and its centre shifts of
$L$ along the $x_1$-axis. The critical curve intersects the $x_1$-axis
in $x_1 \simeq L {\pm} 1$. The changes in the width and in the height of
the critical curve are of order ${\cal O}(L^2)$.

The four extremal points of the critical curve can be mapped onto the
source plane through the lens equation to locate the cusps of the
caustic. We find a diamond-shaped caustic with four cusps, centred in
$\left\{ y_1,y_2 \right\}=\left\{ L(\frac{3}{2}r-1),0 \right\}$. The
axes, of semi-width $\sim L^2$, are parallel to the coordinate axes.
The orientation and the position on the caustic depends on both the
strength and orientation of the angular momentum and on the radius of
the lens. When the source is inside the caustic, two additional images
form. Since the axial symmetry is broken by the gravito-magnetic
field, the Einstein ring is no more produced. A source, which is
inside the central caustic, is imaged in a cross pattern, see
Fig.~\ref{SISCrossGRM}.

\section{Time delay}
\label{dela}

\begin{figure}
        \resizebox{\hsize}{!}{\includegraphics{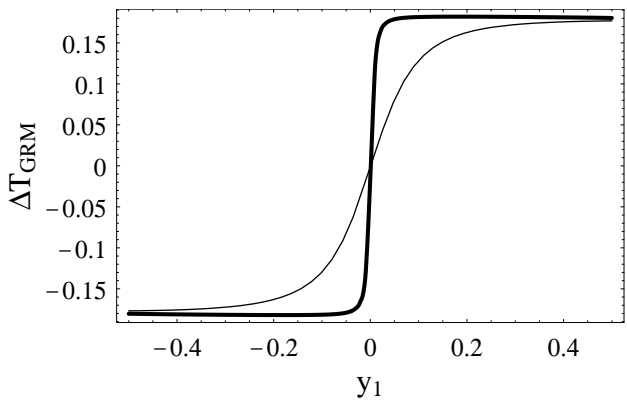}}
        \caption{The gravito-magnetic time delay (in units of days) for a source moving
        parallelly to the $y_1$-axis. Thick and thin lines are for
        $y_2=0.01$ and 0.1, respectively. It is $L=4{\times} 10^{-5}$ and $r=15$.}
        \label{SISDeltaTGRM}
\end{figure}

As we have seen in Section \ref{basi}, the time delay for a spinning
SIS is made of three main contributions: the geometrical time delay,
$\Delta t_{\rm geom}$, the unperturbed gravitational time delay by a
static SIS, $\Delta t_{\rm 0}$, and, finally, the gravito-magnetic
time delay, $\Delta t_\mathrm{GRM}$,
\beq
\lab{dela1}
\Delta t \simeq \Delta t_{\rm geom} + \Delta t_0 + \Delta t_\mathrm{GRM}.
\eeq
The main contribution to the gravitational time delay of a light ray
with respect to the unperturbed path can be re-written as
\beq
\lab{dela2}
c \Delta t_0 = 4 \pi (1+z_\mathrm{d}) D_{\rm d}\left(
\frac{\sigma_\mathrm{v}}{c} \right)^2 \theta,
\eeq
where $\theta$ is the modulus of the angular position in the lens
plane, $\bftheta = \bfxi /D_\mathrm{d}$. The gravito-magnetic time
delay is
\begin{eqnarray}
c \Delta t_{\rm GRM} & \simeq & - 24 \pi \left(
\frac{\sigma_\mathrm{v}}{c}
\right)^3  (1+z_\mathrm{d}) D_{\rm d}
\left( \frac{3}{2}- \frac{\theta}{\theta_\mathrm{SIS}} \right)
\theta_1  \lambda \lab{dela3}
\\
& \simeq & -36 \pi  \left( \frac{\sigma_\mathrm{v}}{c}
\right)^3 (1+z_{\rm d}) D_{\rm d}
 \theta_1 \lambda \lab{dela4}
\end{eqnarray}
where $\theta_1$ is the angular distance of an image from the
projected rotation axis and $\theta_\mathrm{SIS}
\equiv R_\mathrm{SIS}/D_\mathrm{d}$. To obtain equation~(\ref{dela4}),
we have used the relation $x \ll r$. An approximate relation holds
between $\Delta t_0$ and the gravito-magnetic time delay,
\beq
\Delta t_\mathrm{GRM} \simeq -9 \cos \vartheta \left( \frac{\sigma_\mathrm{v}}{c}
\right) \lambda \Delta t_0.
\eeq
Since $L \ll 1$, the angular separation between the two images is nearly
\beq
\Delta \theta \simeq 8 \pi \left( \frac{\sigma_\mathrm{v}}{c}
\right)^2 \frac{D_{\rm ds}}{D_{\rm s}},
\eeq
and the gravito-magnetic induced retardation between the two images,
$\Delta T_\mathrm{GRM} \equiv \Delta t_\mathrm{GRM}(\bfx_a) -\Delta
t_\mathrm{GRM}(\bfx_b)$, can be approximated as
\beq
c \Delta T_{\rm GRM} \simeq 288
\pi^2 \cos \vartheta \frac{D_{\rm ds} D_{\rm d} }{D_{\rm s}}
\left( \frac{\sigma_\mathrm{v}}{c} \right)^5  \lambda.
\eeq
Since the two images are nearly collinear with the lens centre, the
gravito-magnetic time delay is nearly independent of the cut-off
radius.

For a typical lensing galaxy at $z_{\rm d}=0.5$ with
$\sigma_\mathrm{v} \sim 250$~Km~s$^{-1}$, and a background source at
$z_{\rm s} = 2.0$, the gravito-magnetic time delay is $\sim
0.1-0.2$~days for $\lambda \sim 0.05$-0.1. In Fig.~\ref{SISDeltaTGRM},
the time delay between the images is plotted as a function of the
source position for a source moving perpendicularly to the projected
rotation axis. The distance of the source from the $y_1$-axis
determines the width of the transition between the extremal values.

\subsection{The Hubble constant}

\begin{figure}
        \resizebox{\hsize}{!}{\includegraphics{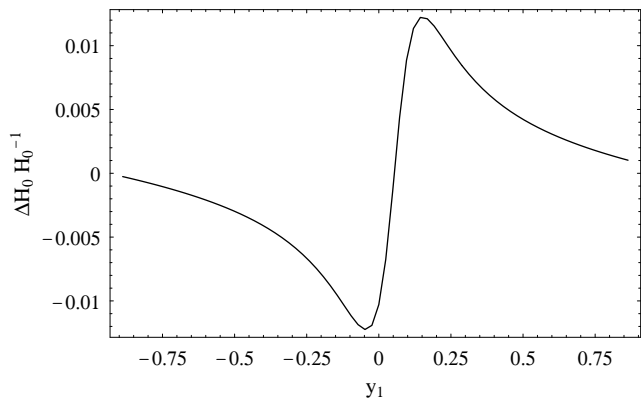}}
        \caption{Relative error in the estimate of the Hubble constant, due to
        neglecting the gravito-magnetic field, for a source moving with fixed $y_2=0.1$.
        It is $L=2.5 {\times} 10^{-3}$ and $r=15$.}
        \label{SISVarH0GRM}
\end{figure}

Any gravitational lensing system can be used to determine the Hubble
constant \citep{ref64}. Neglecting the gravito-magnetic correction can
induce an error in the estimate of the Hubble constant. In general, it
is
\beq
H_0 \Delta t ={\cal F}(\sigma_\mathrm{v},...,z_{\rm d},z_{\rm
s};\Omega_{i0}),
\eeq
where the dimensionless function ${\cal F}$ depends on the lens
parameters and on the cosmological density parameters, but the latter
dependence is not very strong. A lens model which reproduces the
positions and magnifications of the images provides an estimate of the
scaled time delay $H_0 \Delta t$ between the images. Therefore, a
measurement of $\Delta t$ will yield the Hubble constant. Let us
consider a rotating galaxy, described by a SIS with known dispersion
velocity and redshift, which forms multiple images of a background
quasar at redshift $z_{\rm s}$. An observer can measure the time delay
between the two images, $\Delta T_\mathrm{obs}$, and their positions,
$\bfx_a$ and $\bfx_b$. The unknown source position $\bfy$ can be
obtained by inverting the lens equation. In terms of the dimensionless
Fermat potential $\phi$, the measured Hubble constant turns out to be
\beq
H_0 = \frac{1}{\Delta T_\mathrm{obs} }F(z_{\rm d},z_{\rm
s},\sigma_\mathrm{v})|\phi(\bfx_a,\bfy)-
\phi(\bfx_b,\bfy)|,
\eeq
where
\beq
F(z_{\rm d},z_{\rm s},\sigma_\mathrm{v}) \equiv (1+z_{\rm d})\left[
4\pi
\left(
\frac{\sigma_\mathrm{v}}{c}\right)^2\right]^2 \frac{r_{\rm d} r_{\rm ds}}{r_{\rm s}}
\eeq
and $r$ is the angular diameter distance in units of $c/H_0$.

If, we assume a static lens model to model the data, the estimated
Hubble constant is
\beq
H_0^\mathrm{ST}=\frac{1}{\Delta T_\mathrm{obs}}F(z_{\rm d},z_{\rm
s},\sigma_\mathrm{v})2y^\mathrm{ST}.
\eeq
where the ``not correct" estimated position of the source is
\beq
\bfy^{\rm ST}=\frac{1}{2}\left\{ \sum_{a,b}\bfx_i-\frac{\bfx_i}{|\bfx_i|} \right\}.
\eeq
The relative error in the determination of the Hubble constant is
\beq
\frac{\Delta H_0}{H_0}=\frac{2y^{\rm ST}-|\phi(\bfx_a,\bfy)- \phi(\bfx_b,\bfy)|}{|\phi(\bfx_a,\bfy)- \phi(\bfx_b,\bfy)|};
\eeq
For a source at fixed $y_2$, the maximum error is $\sim 1/2\left|L/y_2
\right|$, see Fig.~\ref{SISVarH0GRM}. Since usually $L
\stackrel{<}{\sim} 10^{-4}$, the induced relative error is really
negligible.

\section{Discussion}
\label{disc}

Since the physics of gravitational lensing is well understood, the
gravito-magnetic time delay may provide a new observable for the
determination of the total mass and angular momentum of the lensing
body \citep{ci+ri03}. Although a detailed model may be required to
reproduce the overall mass distribution in the lens, interpretation of
time delay is based on a limited number of parameters. Provided the
cluster where the deflector lies can be described by a simple
expansion, the only parameters needed to model the time delay are
those needed to vary the average surface density of the lens near the
images and to change the ratio between the quadrupole moment of the
lens and the environment \citep{ko+sc04}. Furthermore, the presence of
an observed Einstein ring can provide strong independent constraints
on the mass distribution.

We have developed our treatment of the gravito-magnetic time delay by
modeling the lens as a SIS. Isothermal models are supported by both
theoretical prejudices and estimates from observations of early-type
galaxies so that SIS turns out to be a surprisingly realistic starting
point for modeling lens potentials. Gravito-magnetic time delays of
$\sim 0.1$-0.2 days can be produced in typical lensing systems. A
broad range of methods for reliably determining time delays from
typical data and a deep understanding of the systematic problems has
been developed in last years. Time delay estimates are more and more
accurate and an accuracy of 0.2~days has been already obtained in the
case of B0218+357 \citep{big+al99}, so that the detection of the
gravito-magnetic time delay will be soon within astronomers' reach.
Observations with radio interferometers or $HST$ can measure the
relative positions of the images and lenses to accuracies
$\stackrel{<}{\sim} 0.005\arcsec$. Shifts in the image positions due
to a gravito-magnetic field are usually well above this limit.

Our results are nearly unaffected by the presence of low-mass
satellites and stars. These substructures do not have any impact on
time delays and can only produce random perturbations of approximately
$0.001\arcsec$ in image positions \citep{ko+sc04}, quite below the
gravito-magnetic effect. Deviations from circular symmetry due to
either the ellipticity of the deflector or the local tidal gravity
field from nearby objects should be considered too. However, for a
singular isothermal model with arbitrary structure, the time delays
turn out of be independent of the angular structure \citep{ko+sc04}.
Other higher-order effects, such as the delay due to the quadrupole
moment of the deflector, should be considered in addition to the
gravito-magnetic time delay. Unlike other effects, a gravito-magnetic
field can break the circular symmetry of the lens, inducing
characteristic features in lensing events \citep{ser03prd}. At least
in principle and for some configurations of the images, a suitable
combinations of the observable quantities can be used to remove
additional effects, due to a quadrupole moment, from observational
data \citep{ci+ri02,ciu+al03}. Furthermore, when the lensed images lie
on opposite sides of the lens galaxy, the time delay becomes nearly
insensitive to the quadrupole structure of the deflector
\citep{ko+sc04}.

Measurements of gravito-magnetic time delays could offer an
interesting prospective to address the ``angular momentum problem".
Cold dark matter models of universe with a substantial cosmological
constant appear to fit large scale structure observations well, but
some areas of possible disagreement between theory and observations
still persist. The most serious small scale problem regards the origin
and angular momentum in galaxies \citep{pri04}. Two ``angular momentum
problems" prevent the formation of realistic spiral galaxies in
numerical simulations \citep{pri04}: {\it i)} overcooling in merging
satellites with too much transfer of angular momentum to the dark halo
and {\it ii)} the wrong distribution of specific angular momentum in
halos, if the baryonic material has the same angular momentum
distribution as dark matter halo. Detection of gravito-magnetic
effects in gravitational lensing systems due to the spin of the
deflector, either through time-delays measurements, as discussed in
this paper, or through observations of the rotation of the plane of
polarization of light waves from the background source \citep{ser04},
could provide direct estimates of angular momentum and help in
developing a better understanding of astrophysics in galaxies.

\bibliographystyle{mn2e}
\bibliography{time_delay}

\end{document}